# Developers' Perception of GitHub Actions: A Survey Analysis

Sk Golam Saroar
York University
Toronto, Ontario, Canada
saroar@yorku.ca

Maleknaz Nayebi
York University
Toronto, Ontario, Canada
mnayebi@yorku.ca

## ABSTRACT

GitHub introduced "Actions" in 2019 to increase workflow velocity and add customized automation to the repositories. Any individual can develop Actions for automating workflow on GitHub repositories and others can reuse them whenever open source. GitHub introduced its marketplace for commercializing and sharing these automation tools, which currently hosts 16,730 Actions. Yet, there are numerous Actions that are developed and distributed in local repositories and outside the Marketplace. So far, the research community conducted mining studies to understand Actions with a significant focus on CI/CD. We conducted a survey study with 90 Action developers and users to understand the motivations and best practices in using, developing, and debugging Actions, and the challenges associated with these tasks. We found that developers prefer Actions with verified creators and more stars when choosing between similar Actions, and often switch to an alternative Action when facing bugs or a lack of documentation. 60.87% of the developers consider the composition of YAML files, which are essential for Action integration, challenging and error-prone. They mainly check Q&A forums to fix issues with these YAML files. Finally, developers tend to avoid using Actions (and hence automation) to reduce complexity, and security risk, or when the benefits of Actions are not worth the cost/effort of setting up Actions for automation.

## KEYWORDS

GitHub Actions, open source, workflow automation, software engineering, survey, developer's perception

## 1 INTRODUCTION

To increase workflow velocity, GitHub introduced GitHub Actions which brings automation directly into the software development lifecycle on GitHub. GitHub Actions are built around GitHub events and are triggered by developers' activities, such as creating a pull request or opening a new branch on a repository [6]. Open-source software development thrives on collaboration. GitHub, the most prominent social coding platform, provides features such as issue tracking and pull requests to facilitate open-source software development. This results in an increased workload for developers and the subsequent use of automation tools to perform various repetitive tasks. GitHub Actions (sometimes we refer to these as Actions) are the main leverage in the hand of software developers to automate their tasks.

At present, there are 16,730 Actions on the GitHub Marketplace[1] across 21 Categories [27]. There is a growing trajectory in the number of offered Actions on the marketplace in a way that the number of Actions in the marketplace in December 2022 was 2.1 times more than the number of Actions in April 2021 (16,730 compared to 7,878 Actions). Even further, not all actions are distributed on the GitHub marketplace, and many developers design and keep their Actions within local repositories and never put them on the marketplace. Decan et al. [9] analyzed the most frequent automation practices on GitHub and found that 43.9% of repositories in their dataset adopt and use Actions for automating parts of their workflows.

In this environment, for any given task, multiple competing Actions can be expected. This raises a new challenge for the practitioners to decide which Action to use [15]. GitHub Actions are created *by* the developers *for* the developers. These Actions reflect a real-world effort for task automation in software teams. It is unclear how developers look up these Actions or what motivates them to choose a particular Action over the others. Even further, while GitHub officially provides a platform for distributing Actions, not every developer chooses to distribute their actions on this marketplace. In addition, it is important to study "why" developers create new Actions despite having multiple alternatives to adopt or reuse. Further, to our knowledge, no research distinguishes developers who create and contribute to Actions (providers) from developers who only use Actions (users). We are the first to identify these different types of Action adopters to understand their motivation for developing and using (or not using) Actions and the varying challenges associated with each of these tasks.

So far, the software engineering mining community investigated the usage and adoption of GitHub Actions but not Action development. A number of mining studies evaluate the basic adoption and usage details of GitHub Actions. These studies retrieve a random set of repositories that use and adopt Actions and study a particular automation task. For instance, Golzadeh et al. [12] focused on continuous integration (CI) and mined 91,810 GitHub repositories to analyze how the CI landscape has changed since GitHub Action was introduced. They reported that the introduction of Actions aligns with a decreasing growth rate of CI usage for Travis, CircleCI, and Azure. Kinsman et al. [15] analyzed 3,190 repositories to investigate how several development activity indicators change after Action adoption. In our study, to understand developers' perceptions, we conducted a survey study [16] involving a total of 90 software developers on GitHub. The goals of our study are to understand (i) why and how developers publish or use GitHub Actions, and (ii) what challenges developers face in creating, publishing, or using Actions. Through an exploratory analysis [35] of the survey responses, we answer the following research questions:

---
[1]https://github.com/marketplace?type=actions

Table 1: Comparison between existing literature and our work

| Existing study | Contributions | Comparison |
|---|---|---|
| Kinsman et al. [15], Wessel et al. [33] | Mined GitHub repositories to understand the impact of Actions on the pull-request process. To understand developers' perceptions of Actions, they analyzed comments on GitHub issues that discuss Actions. | We focus on a qualitative study and survey developers to better understand their perceptions of Actions. For example, while some developers discuss issues in the Action repository, our survey data shows that most developers use Q/A forums such as Stack Overflow to discuss issues regarding Actions. |
| Chen et al. [7] | Also mined GitHub repositories to understand the impact of Actions on the commit frequency, pull request, and issue resolution efficiency. To investigate how developers configure their Actions, they performed an analysis of the YAML files used in Action repositories. | While a statistical analysis such as [7] gives a useful overview of developer practices in writing YAML files, it is important to understand how they write these files, what issues they face, and how they debug these issues. In our study, we survey practitioners on these topics and present our findings. |
| Decan et al. [9] | Analyzed the characteristics of the repositories that use GitHub Actions to study the automation practices, which workflows are automated, and which actions are reused. | Instead of studying repositories, we study developers to understand their motivation, decision criteria, and challenges in developing and using Actions. We ask developers when they automate a task, or why they do not reuse Actions. |
| Golzadeh et al. [12] | Mined GitHub repositories to analyze how the CI landscape has changed since Actions was introduced. | Although we asked developers what other automation services they use, we did not ask them if they migrated to Actions from other CI services. |
| Valenzuela-Toledo et al. [29] | Mined commits involving YAML files to highlight deficiencies in the way Actions workflows are produced and maintained. | We asked practitioners about their perceived challenges in writing YAML. |

**RQ1-** What motivates software developers to create, publish, and use GitHub Actions?
*What and How:* We are interested to understand why developers create Actions, whether or not they study existing Actions before creating their own, why they publish Actions on the marketplace, and which tasks they automate using Actions. Therefore, we surveyed participants with two close-ended and two open-ended questions.

**RQ2-** What are the decision criteria for developers to create and use GitHub Actions?
*What and How:* After asking developers why they create and use Actions, we move on to explore how they do it. We surveyed participants to learn when they automate tasks (and when they do not), how they find Actions, choose between similar Actions, and configure and debug issues with their workflow files. To this end, we had a total of eight questions, one of which was open-ended.

**RQ3-** What challenges do developers face in creating, publishing, and using Actions?
*What and How:* We analyzed the challenges in creating Actions, publishing them to the marketplace, as well as adopting and configuring the workflow files when using the Actions. We also asked participants about the issues in workflow automation and the GitHub marketplace in general. We asked nine questions, five of which were open-ended.

This study is the first to distinguish different kinds of Action adopters and try to understand their perceptions of GitHub Actions. We designed a survey and collected data involving the responses of 90 software engineers. We provide a detailed analysis of the survey responses that give insights on Action providers and users- their motivation, decision criteria, and challenges in creating, publishing, and using Actions. Our findings can be used to help users choose the right Actions for their development tasks and help providers identify areas of improvement. This should also help researchers understand the state of practice in software automation so that they can conduct relevant research to advance the community.

Section 2 presents the related work and Section 3 describes survey design, participant selection, and data analysis. Section 4 presents the results of the survey analysis. Section 5 discusses the implications and recommendations. Section 6 addresses the threats to validity and Section 7 concludes this paper.

## 2 RELATED WORK

Kinsman et al. [15, 33] quantitatively analyzed the impact of adopting GitHub Actions, particularly focusing on how the dynamics of pull requests changed for 3,190 GitHub projects after they adopted Actions. Their results indicate that the adoption of Actions increases the number of rejected pull requests and decreases the number of commits in merged pull requests. In addition, the authors stated that adopting Actions led to requiring more time to accept a pull request. Similarly in [7], Chen et al. investigated the impact of Actions on the commit frequency, pull request, and issue resolution efficiency. They found that after adopting Actions, commit frequency decreases, the number of closed issues increases, and the number of pull requests decreases. However, unlike [33], the authors found that adopting Actions reduced the pull-request resolution latency. Decan et al. [9] analyzed the characteristics of the repositories that used GitHub Actions, which workflows were automated using GitHub Actions, as well as the most frequent automation practices on GitHub. Golzadeh et al. [12] focused on continuous integration (CI) and analyzed how the CI landscape has changed since GitHub Action was introduced. As the authors studied how frequently CIs are being replaced by an alternative, they found that GitHub Actions attracted most of the migrations and took only 18 Months to overtake Travis CI in popularity. Even so, while there have been numerous research on the use [13, 14, 30], misuse [11, 31, 38], and areas of improvement [10, 34] for continuous integration, there is a substantial lack of similar research on GitHub Actions. Valenzuela-Toledo et al. [29] investigated the types of GitHub Actions workflows modifications performed by developers in 10 Popular GitHub repositories. They manually inspected 222 Commits related to workflow changes and identified 11 Different types of workflow modifications. The authors highlighted the need for adequate tooling to support refactoring, debugging, and code editing of Actions workflows. Calefato et al. [5] focused on GitHub Actions and CML[2] to present a preliminary analysis of the MLOps practices in GitHub. They studied 397 Actions workflows and 38 CML workflows to report whether workflow automation was prevalent on GitHub, what events were used to trigger these workflows, and what tasks were most frequently automated. The authors concluded that there were very few open-source projects involving ML-enabled components or systems that used GitHub Actions. Benedetti et al. [2] investigated the security issues affecting Action workflows and implemented a methodology (GHAST) for automatically assessing the presence of security issues in workflows. GHAST was used against 50 open-source projects to provide an overview of the current security landscape of Action workflows. Koishybayev et al. [21] created the workflow auditing GitHub Action, GWChecker, to help mitigate common security errors in the YAML configuration for CI/CD workflows.

Table 2: Questions in the survey and their recipients (P = Provider, U = User, NU = Non-user). Subscript with a question number shows the number of responses.

| RQ1 | Rcpt | What motivates developers to create, publish, and use GitHub Actions? |
|---|---|---|
| $1_{44}$ | P | Can you specify why you did not reuse an existing Action but created one yourself? *(four options and others)* |
| $2_{44}$ | P | When there are similar Actions (as yours) created by others, to what extent do you look into their functionality and scope? *(Likert scale)* |
| $3_{44}$ | P | In your opinion, what are the advantages of publishing to GitHub Marketplace (if any)? *(textbox)* |
| $4_{69}$ | P, U | What is/are the primary task(s) you are using GitHub Actions for? (mention up to three tasks) *(textbox)* |
| **RQ2** | | **What are the decision criteria for developers to create and use GitHub Actions?** |
| $5_{69}$ | P, U | How do you identify if a task should be automated in your GitHub workflow? (select all that apply) *(three options and others)* |
| $6_{69}$ | P, U | When having a task requiring automation, what is your primary way of identifying a proper Action for use? (select all that apply) *(five options and others)* |
| $7_{69}$ | P, U | When multiple Actions are available for a task, you choose the Action having __? (please drag the options to order them from highest to lowest preference according to you) *(six options and others)* |
| $8_{69}$ | P, U | In the past, have you switched to using a different Action for your tasks? If yes, what was the reason for switching? (select all that apply) *(seven options and others)* |
| $9_{69}$ | P, U | How do you write your workflow (YAML) files? (select all that apply) *(four options and others)* |
| $10_{44}$ | P | Are you using self-hosted runners? *(yes/no)* |
| $11_{44}$ | P | Why are you using self-hosted runners? *(five options and others)* |
| $12_{69}$ | P, U | Have you chosen not to automate certain tasks even though GitHub Actions and other tools are available? If yes, please briefly specify which tasks and why. *(textbox)* |
| **RQ3** | | **What challenges do developers face in creating, publishing, and using Actions?** |
| $13_{44}$ | P | What challenges do you face in developing your Action(s) on GitHub (if any)? *(textbox)* |
| $14_{44}$ | P | Have you published your Action on the GitHub Marketplace? If not, please specify why. *(four options and others)* |
| $15_{69}$ | P, U | What challenges do you face in configuring your workflow (YAML) files (if any)? *(textbox)* |
| $16_{25}$ | U | If facing an issue in configuring your YAML file, where do you primarily look for information? (select all that apply) *(four options and others)* |
| $17_{69}$ | P, U | What challenges have you experienced in automating workflows using GitHub Actions (i.e. are there certain tasks that you could not automate)? *(textbox)* |
| $18_{25}$ | U | GitHub Marketplace helps in discovering tools to improve workflows. Have you faced any challanges in using the marketplace? *(textbox)* |
| | | **Why do developers not use GitHub Actions?** |
| $19_{21}$ | NU | Have you heard about GitHub Actions for automating workflows in GitHub? *(yes/no)* |
| $20_{21}$ | NU | Why do you not use GitHub Actions for automating your workflows? *(textbox)* |
| $21_{46}$ | U, NU | Which other workflow automation services have you used for your GitHub repositories? (select all that apply) *(seven options and others)* |
| $22_{25}$ | U | Do you believe that GitHub Actions have helped you in your role as a software engineer? *(5-point Likert scale)* |

Table 1 shows an overview of the prior studies related to Actions and how our work compares with them. All the studies rely on mining GitHub repositories and performing quantitative analysis for retrieving conclusions. To the best of our knowledge, this is the first study seeking developers' perceptions of the benefits and drawbacks of creating, publishing, and using GitHub Actions.

## 3 METHODS

Following the empirical guidelines for survey design in software engineering [20], we discuss our survey design, participant selection, and analysis of the survey responses. The survey is fully reviewed and approved by the ethics board at [...] number [...][3]. We followed the established empirical protocols for performing surveys following Kitchenham and Pfleeger [16]. In consideration of the current status quo of the state of the art and the nature of the mining studies on the repositories, we decided to focus this survey research on the *diagnostic evaluation* [35] to understand "*why*" and "*how*" of developers decision in creating and using GitHub Actions. Figure 1 shows the steps in our survey method.

---

[2] https://cml.dev/

[3] This information is omitted due to double-blind review guidelines

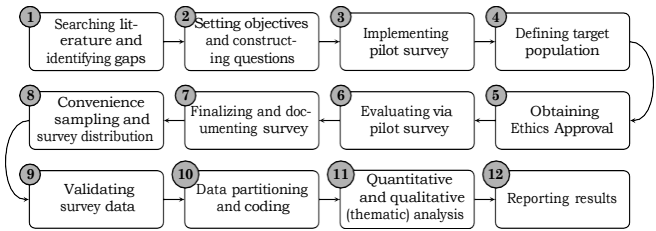

**Figure 1: The stages of our survey method**

### 3.1 Survey Design

We started by setting the objectives of the survey and performing a literature study. The objectives define the survey context and identify what areas to cover and what types of information are needed [20]. We identified three objectives for this survey to address the gaps in the mining studies and to better understand developers;

(1) understanding *why* developers create, publish, and use Actions;
(2) finding out *when* and *how* developers create, search for, switch between, and configure Actions,
(3) identifying the *challenges* developers face in creating and using Actions as well as using the GitHub Marketplace.

Having the objectives, we brainstormed on the potential questions for each objective. We filtered our questions based on similarity/redundancy and finalized 22 unique questions that were aligned with our objectives. We performed a second round of literature review to identify any relevant research and ensure that the questions are novel. We identified three types of developers based on their familiarity with GitHub Actions. Some developers created and published **(P)** Actions, and are highly familiar with the design, adoption, and maintenance decisions. However, not all creators are publishers as they might only use it within a limited context of their own repository. We will refer to this group as Providers. The second group has mostly adopted and used Actions to automate tasks in their repository, we refer to them as Users **(U)**. The third group, Non-users **(NU)**, are the least engaged users as they never used, and in some cases, never heard of GitHub Actions and hence not helpful in achieving our survey goals. To answer our research questions effectively, we curated our questionnaire for each of these groups. In Table 2, we list all the questions of the survey separated per objective and type of developers group (P, U, and NU).

Overall, the survey included three demographic questions, nine multiple-choice, two yes/no questions, two Likert-scale questions, one ranking question, and eight open-ended questions. The demographic questions concern the participants' experience in software development (i.e, less than 1 year, more than 10 years, etc.), their role in their company (i.e, software developer, project manager, other), and their self-perceived level of familiarity with GitHub Actions (i.e created Actions, only used Actions, never used Actions, etc). To curate the survey based on the participants' familiarity and group type, for example, not asking developers what challenges they face in developing Actions while they have never developed or even heard of Actions, we used our third demographic question to design our survey in a way that participants were asked only the questions that applied to them. To build the survey instrument, we used `Qualtrics`. Our survey was anonymous, and we did not collect any identifying information about the participants.

### 3.2 Selecting participants

We invited 10 Developers from our close connections for a pilot study and evaluate our survey, i.e. making sure the questions were understandable, and the survey was *reliable* and *valid* [18]. A survey is reliable if we administer it many times and get roughly the same distribution of results each time. We employed the *alternative form reliability* where we asked the invitees to participate twice but reworded a few questions the second time. On the other hand, *validity* is concerned with how well a survey measures its' objectives. We used *content validity* where we asked invitees to provide us with written feedback (if any) regarding the survey. We also asked them to record how many minutes it took them to complete the survey. We analyzed the pilot responses to assess if we needed to change anything about the survey before we sent it out to a larger pool of participants. Based on the feedback from our initial participants, we revised two questions to eliminate ambiguity and set 15 minutes as the expected time to complete the survey. We recruited participants for the survey by sending personalized emails to GitHub developers (random sampling), and by advertising the survey on our social media (convenience sampling) [17]. We randomly sampled 250 developers who created or contributed to developing Actions and used our academic email to send personalized email invitations. We also posted the survey on our social media platforms (Twitter and LinkedIn). We received 175 clicks on our survey.

### 3.3 Analyzing the data

As the first step of our data analysis, we checked the completeness of our responses. Although we received 175 clicks on our survey, upon close inspection, we found 74 responses as either entirely empty responses or only including responses to the demographic questions. An additional 11 responses were 30% to 70% complete. For the sake of completeness, we left out these responses too. Hence, we excluded a total of 85 responses from our analysis. We then assigned an ID to each response and performed our survey data analysis on the 90 fully-completed responses. However, participants only received and responded to the questions that applied to them (based on a demographic question, see section 3.1). For this reason, we had different sample sizes across questions.

We employed statistical and qualitative approaches to analyze the survey responses. We performed a thematic analysis for the qualitative open-ended questions. The thematic analysis involves reading through a data set and looking for patterns in meaning in order to derive themes within the context of the research question [4, 8]. In the initial iteration of the analysis, two authors of the paper independently annotated all the responses to the open-ended questions. After that, they compared the categories and discussed any ambiguities or disagreements. For the quantitative analysis, we performed statistical tests whenever applicable.

## 4 RESULTS

In this section, we present the results by first describing the survey participants based on demographic information. Then, we present the motivation, decision criteria, and perceived challenges faced by both users and providers of GitHub Actions.

### 4.1 Survey Participants

We asked our participants three demographic questions- their experience in years, their role within their companies, and their familiarity with GitHub Actions.

**Experience and Role$_{(90)}$**: 33.33% of our survey participants have between one to two years of experience in the software industry. This is followed by 25.56% participants with experience between three to six years. 17.78% of participants have less than one year of experience while 16.67% are with more than 10 years of experience. The smallest group consists of 6.67% participants who have an experience between seven to 10 years. These participants vary in their roles in a team. 64.44% (58 among 90) of the participants are *developers* followed by *managers* (14.45%) and *DevOps engineers* (8.89%). We also received responses from *Consultants* (4.44%), *Systems Engineers* (3.33%), *Testers* (2.22%), and *Security Analysts* (2.22%).

**Type$_{(90)}$**: We partitioned the responses following Kitchenham and Pfleegar [19] and considered three types of GitHub users.

  Provider (P): Participants who have created Actions or contributed to Actions or both. We assume this group has also used Actions.

  User (U): Participants who have neither created nor contributed to Actions but used Actions for one or more tasks.

  Non-user (NU): Participants who have never created, contributed to or used Actions.

54.44% of our survey participants are *providers*. Among these participants, five have only contributed to one or more Action repositories (contributor) but never created an Action. 22.22% (20 out of 90) of our respondents are *users* and the remaining 21 (23.33%) participants are *non-users*. 15 out of these 21 participants said that they have not heard about GitHub Actions.

### 4.2 RQ1 - Developers' motivation for creating, publishing, and reusing Actions

We asked four questions (Q1 to Q4 in Table 2) to evaluate what motivates developers to create Actions, publish them on the marketplace, and use them for automation tasks on GitHub.

**Creating Actions instead of reusing$_{(44)}$**: While there are over 15,000 Actions in existence, an increasing number of Actions are published every day on the Marketplace. This creates the question as to *why* developers prefer to develop new Actions rather than reusing the existing ones. We intend to identify the existing gaps in the current practice. In response, 56.82% of developers in our survey have stated that they create new Actions mainly because their *requirements are not met* in any of the existing Actions. One-fourth of them believe their requirement is unique and there is *no existing Action* or that *existing Actions have limited functionality* for their purposes. Developers also create Actions when the *existing*

**Table 3: Motivations and challenges for creating Actions (P$_{44}$)**

| Reason for creating Action | % of participants |
|---|---|
| No existing Action was available. | 56.82 |
| Existing actions are limited in functionality. | 25 |
| Existing actions are not performant enough. | 11.36 |
| Not sure how to find and reuse Actions. | 4.55 |
| Existing actions were too complex. | 2.27 |
| **Check out similar Actions first?** | **% of participants** |
| In great detail | 45.45 |
| Somewhat detail | 43.18 |
| Casually | 11.36 |
| **Using self-hosted runners?** | **No. of participants** |
| No | 23 |
| Not sure | 5 |
| Yes | **16** |
|     Use cloud/local machines already paid for | 4 |
|     Create custom hardware configurations | 4 |
|     Install software from local network | 3 |
|     Specific security requirements | 3 |
|     OS not offered by GitHub-hosted runners | 2 |
| **Challenges in creating Actions** | **% of participants** |
| Test-ability | 25 |
| Limited resources | 18.18 |
| Privacy and security | 6.82 |
| Integration | 6.82 |
| Other | 18.18 |
| No challenges | 25 |

*Actions' performance was not good enough* (11.36%) or they are *not sure how to find and reuse Actions* (4.55%) (detailed in Table 3).

> "The action was too complex for my task and quite unreasonable in design, so I built mine. [*Survey Participant-98*]"

We know from the Marketplace trend that developers often create new Actions instead of reusing existing Actions. We are interested to understand how and to what extent the functionality and scope of similar Actions created by others impact their design decisions. Table 3 shows that 45.55% of developers stated that they look into others' Actions *in great detail*. 43.18% of developers said they checked similar Actions in a *somewhat detailed* fashion. Only 11.36% of developers checked similar Actions *casually*.

**Table 4: Perceived Advantages of publishing Actions to the marketplace and why developers do not publish (P$_{44}$)**

| Advantages to Publishing | % of participants |
|---|---|
| Visibility | 27.27 |
| Attracting contributions | 27.27 |
| Reusability | 25 |
| Accessibility | 20.45 |
| Improved code quality | 11.36 |
| Other (Get feedback, make connections) | 9.09 |
| **Published Action?** | **No. of participants** |
| Yes | 19 |
| No | **25** |
|     Takes too much time and effort | 9 |
|     Does not have permission to publish | 9 |
|     Does not see benefit in publishing | 5 |
|     Does not know how to publish | 2 |

**Publishing Actions to GitHub marketplace**$_{(44)}$: There are currently 16,730 Actions on the marketplace across 21 categories. Developers in our survey stated that publishing Actions comes with many advantages such that Actions are easily *discoverable* (27.27%), *attract contributions* (27.27%), and *become reusable* (25%). Open-source collaboration often results in *improved code quality*, which is another reason (according to 11.36% participants) for publishing Actions (see Table 4).

> *Publishing to the marketplace increases an organization's visibility and attracts prospective clients/talents. It also helps an individual in building their resume. Besides, going open source can greatly improve the quality of code. [S-19]*

> *Re-usability, and reduction of effort in creating new, similar workflows. [S-80]*

However, when asked if they published their Actions on the Marketplace, the majority (56.82%) of the responses were negative due to various reasons (see section 4.4).

**Primary tasks that are being automated**$_{(69)}$: Most developers (73.91%) are integrating GitHub Actions in their workflows for more than one task. While Actions were originally introduced by GitHub for automating CI/CD tasks within the repositories [7, 12], there is a shift in the usage direction. Only two developers explicitly mentioned *CI/CD* among their automated primary tasks. Rather, developers use Actions to automate Testing (28), Build (26), Deployment (22), and Release (19) - essentially creating their CI/CD pipeline individually and by plugging together different products. Developers also use Actions for *code review*, *linting*, and *benchmarking* (see Figure 2).

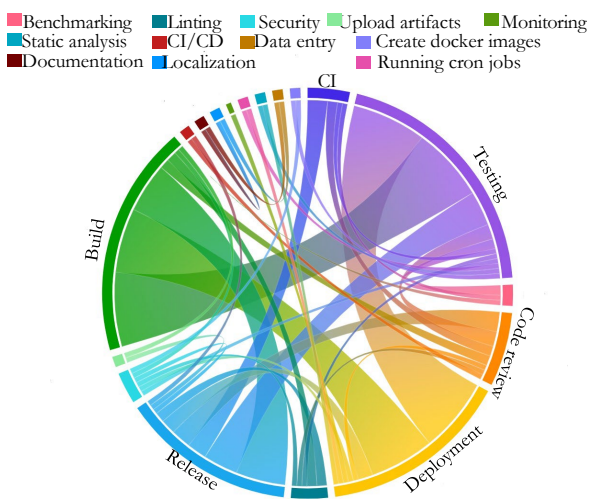

**Figure 2: Tasks developers automate using GitHub Actions.**

**Table 5: Debugging workflow (YAML) files and using GitHub Marketplace (U$_{25}$)**

| Debug YAMLs by | % of participants |
|---|---|
| Check Q/A forums such as Stack Overflow | 92 |
| Create a pull request/bug report for Action | 36 |
| Contact the Action developers | 28 |
| Ask on social media such as Reddit/Twitter | 24 |
| **Problem with marketplace** | **% of participants** |
| No challenges | 60 |
| Searching for products | 16 |
| Hard to quality check | 8 |
| Less marketed | 4 |
| Does not use Marketplace | 12 |

> Unmet functional (for 56.82%) and non-functional (for 36.56%) requirements were the most frequently mentioned motivation of participants to create Actions. Testing has been the most frequently mentioned task that developers primarily automate using Actions. Getting more visibility and attracting contributions for better quality is developers' main motivation to publish their Actions on the GitHub Marketplace.

### 4.3 RQ2 - Decision criteria for developers to create and use Actions

Knowing why developers create and publish Actions, we now discuss *how* they do it.

**When do users decide to automate tasks**$_{(69)}$: The majority of the Action users (P+U) tend to automate a task in their GitHub workflow *if the task is repetitive* (95.65%) or *if the task has been automated in other repositories* (47.83%). We also observed that 33.33% of the participants automate tasks if it *requires high precision*. Our survey participants also tend to introduce automation in their workflow if a task is *mission critical*, *boring*, or *takes too long to do manually* (see Table 6).

> *We try to automate anything mission-critical, where there is no room for human error. [S-32]*

**Finding Actions for use**$_{(69)}$: Knowing that 39.55% of Actions have never been published on the Marketplace [15] but are repetitively used, we are interested to identify the channels where participants identify a proper Action for their use case (Table 6). 68.12% of the users stated that they find Actions by *internet search, YouTube, forums such as Stack Overflow*, followed by 62.32% users who *browse GitHub Marketplace* to find Actions. Surprisingly, developers are more likely to *create their own Action* (39.13%) when they need one rather than *browsing repository-specific suggestions by GitHub* (37.68%) or even relying on *word of mouth from known connections* (20.29%). We observed that the least experienced developers (with less than a year of experience) are most likely to create their own Actions instead of searching for reusable Actions. While not a popular choice, 2.9% of users have said that they *check other repositories that have automated similar tasks*.

**Choosing the right Action**$_{(69)}$: At the time of writing this paper, there are 16,730 Actions on the GitHub Marketplace across 21

**Table 6: How developers decide, find, and use Actions and the challenges they encounter (P+U$_{69}$)**

| Tasks developers automate | % of participants |
|---|---|
| Repetitive | 95.65 |
| Automated in other repositories | 47.83 |
| Requires high precision | 33.33 |
| Mission critical | 4.35 |
| Not requires thinking | 2.9 |
| Takes long to do manually | 2.9 |
| **How developers find Actions** | **% of participants** |
| Web, YouTube, social media, & forums | 68.12 |
| Browsing GitHub Marketplace | 62.32 |
| Creating own Action | 39.13 |
| Browsing GitHub's suggestions | 37.68 |
| Word of mouth from known connections | 20.29 |
| Checking other repositories | 2.9 |
| **Reason to switch between Actions** | **% of participants** |
| Found an Action of better quality | 33.33 |
| Created own action | 26.09 |
| Previous Action had bugs | 26.09 |
| Previous Action lacks documentation | 23.19 |
| Found an Action with more features | 21.74 |
| Found Actions with more compatibility | 2.9 |
| Never switched between Actions | 40.58 |
| **How developers write YAMLs** | **% of participants** |
| Modifying a previous workflow file | 69.57 |
| Using configurable templates by GitHub | 50.72 |
| Writing from scratch | 49.28 |
| Copying from websites or forums | 43.48 |
| Action documentation | 1.45 |
| **Challenges in writing YAMLs** | **% of participants** |
| No challenges | 39.13 |
| Never configured YAML files | 10.14 |
| Syntax | 11.59 |
| Debugging | 8.7 |
| Generalizability | 7.25 |
| Indentation | 5.8 |
| Missing Intellisense | 4.35 |
| Inadequate documentation | 4.35 |
| Learning curve | 2.9 |
| Other | 5.8 |
| **Challenge in automating workflows** | **% of participants** |
| No challenges | 40.58 |
| Testing & Debugging | 14.49 |
| Complexity | 8.7 |
| Inadequate documentation | 7.25 |
| API limitations | 4.35 |
| Security concerns | 4.35 |
| Lack of support | 2.9 |
| Managing artifacts | 2.9 |
| Integration | 2.9 |
| Other | 8.7 |
| **Prefer not to automate certain tasks?** | **No. of participants** |
| No | 55 |
| Yes | **14** |
|     Preferring manual work | 3 |
|     Reducing complexity | 3 |
|     Cost/effort effectiveness | 5 |
|     Security issue | 1 |
|     Other | 2 |

Categories while many others are being used but not published on the platform [15]. In such a fiercely competitive ecosystem, it is a challenge to find a proper Action. We asked our survey participants how they chose an Action when multiple Actions are available for a particular task. Figure 3 shows the factors and their priorities to our participants. Actions having *verified creator* often have an edge over competitors. 53.62% of participants chose this factor as their first or second preference when choosing between competing Actions. This is closely followed by factors such as *more stars* and *more contributors*, each of which are chosen by 20.29% participants as their first preference. Only 21.74% participants have chosen *published on the marketplace* among their top four preferences. In fact, this factor is the most popular fifth and sixth preference among Action users. *Other* factors mentioned by the participants involved the quality of the actions and ease of code comprehension.

We studied all the results based on the varied experience groups of our participants. However, these results were rarely significantly different between the groups. 33.33% of the most experienced developers (10+ Years) chose *word of mouth* as their first preference in contrast to 15.94% developers choosing this across all groups. On the other hand, 50% of the least experienced developers chose *better description* as their first preference compared to 13.03% developers choosing this criterion overall.

**Switching to a different Action$_{(69)}$**: In this competitive ecosystem, we were interested to learn the frequency and users' decision criteria for switching to alternative Actions. 40.58% (28 of 69) of our respondents stated they never switched between Actions. This can be due to the lack of experience as 26 of these respondents have less than seven years of experience. The others have switched to a different Action because they found an Action with *better quality* (33.33%), *more features* (21.74%), or *better compatibility i.e. regular updates* (2.9%). Users also switch to a different Action when their previous Action *have bugs* (26.09%) or *not properly documented* (23.19%). Interestingly, churningout due to in-house development of Actions was the second-most popular reason for churning out (Table 6). Developers with 10+ years of experience most frequently switch Actions for this reason.

> "Another action kept providing updates while the original action lost its compatibility. [S-69]"

> "Action stopped supporting my required use case. [S-32]"

> "We decided to automate our code review and my senior developer had a suggestion for a code review action which was reviewing and merging pull requests in one.[S-98]"

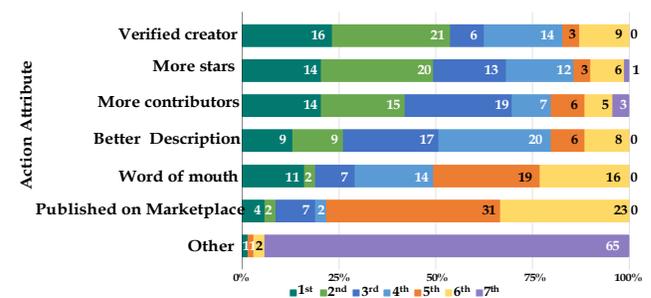

**Figure 3: Developer preferences for choosing an Action when there are multiple Actions to choose from.**

**Writing workflow (YAML) files$_{(69)}$**: Users add a YAML script to their repository for adopting and using an Action and configuring automation for their workflows. The success in adoption and proper use of any Action depends on these files. Among our participants, *modifying a previously used workflow file* (69.57%), and *using configurable templates by GitHub* (50.72%) are the most popular methods for writing such files. Other than that, oftentimes users *write from scratch* (49.28%) or *copy from websites or forums such as Stack Overflow* (43.48%) (see Table 6).

**GitHub-hosted vs self-hosted runners$_{(44)}$**: GitHub Actions are dependent on runners to execute Action workflows. One can host their own runner and customize the environment for Action workflows (self-hosted runners) or use GitHub virtual machines (GitHub-hosted runners). 36.36% (16 out of 44) Providers (P) in the survey stated that they use self-hosted runners (see Table 3).

**Why developers do not automate certain tasks$_{(69)}$**: To have a better picture of the developers' decision for automating development tasks, we asked users whether they prefer *not* to use Actions at times. Only 14 out of 69 (20.28%) respondents preferred not to automate certain tasks. Half of these developers (7 out of 14) had more than 10 years of experience. Developers prefer not to automate tasks when the task is not worth the cost/effort that goes into using Actions (5), or when using Actions increases the complexity of their projects (3), poses security risks (1), poor documentation (2), or simply because they prefer to perform certain tasks manually (3) (Table 6).

> *[I don't use Actions] when repositories are sensitive and I don't want third parties having access. [S-76]*

> The abandoned number of alternative Actions provides software developers with multiple options for automating their tasks and the majority (79.71%) wish to automate all the development tasks. Our study shows that most often the developers tend to automate the repetitive tasks within their software repositories and then either use online (web) search or the GitHub marketplace to find an Action. In this context, the unfulfilled non-functional requirements (65.32%) and the development of in-house solutions (26.09%) motivate developers to migrate from one Action to another.

### 4.4 RQ3- Challenges for developers in creating, publishing, and using Actions

Further, we investigated developers' challenges in creating, publishing, and configuring Actions.

**Creating Actions$_{(44)}$**: 75% of participating developers faced one or more challenges in developing Actions. The most frequently mentioned challenge was the *testing ability* of Actions where 25% of the developers faced difficulty to test and debug their Actions (Table 3). According to 18.18% participants, *limited resources* i.e. lack of documentation and support makes onboarding difficult for new developers. Developers also have *privacy and security concerns* (6.82%) as well as issues with *integrating Actions into workflows* (6.82%). Developers also pointed to the lack of proper monetization channels, the poor design of the Actions framework, limited GitHub APIs, and slow updating of GitHub runners (where Actions run) as challenges in creating Actions.

> *GitHub Actions fail too frequently (especially with Windows runners) with no clear explanation, and these errors are frequently not reproducible. Additionally, GitHub is too slow with adding new macOS versions to runners. [S-71]*

**Publishing to the Marketplace$_{(44)}$**: Not all Actions are available on the marketplace. The majority, 56.82% (25 out of 44), of the Action providers (P) in our survey did not publish their product on the Marketplace. Nine out of these 25 developers reasoned that *publishing needs much time/effort* while other nine developers *did not have permission to publish*. Quite interestingly, the remaining five (out of 25) developers *did not see the benefit in publishing* while two said they *did not know how to publish* (Table 4).

**Configuring workflow (YAML) files$_{(69)}$**: Configuring a workflow using a YAML file is not a trivial task. 50.73% of Action users reported challenges in configuring their workflow files. 11.59% of the developers have said that YAML file's *syntax is confusing and error-prone* and 8.7% developers have commented that YAML files are *difficult to test and debug*. Another 5.8% of the participants *struggle with indentation rules*. This is while others were challenged with the *missing intellisense* (i.e. code completion, syntax highlighting) and *inadequate documentation* about how to configure Actions using YAML (Table 6).

> *I am very used to having intellisense when I write code with any language, so it was a bit challenging to write YAML without intellisense help on GitHub. [S-11]*

> *They're not fully testable locally so if you ever did something wrong, you've to rebase changes over & over again. [S-13]*

> *YAML is a bad for writing configuration code. It's untyped, which frequently leads to serious bugs in configuration code. I wish there was a statically-typed and more reliable alternative to YAML available and officially supported by GitHub Actions. [S-71]*

> *YAML is great to read, and a nightmare to write. [S-72]*

We asked survey participants where they look for information when facing an issue with YAML configurations. 92% of the participants use *Q/A forums such as Stack Overflow* while 36% users *create a pull request/bug report in the Action repository* (Table 5).

**Automating workflows using Actions$_{(69)}$**: 59.42% of users experienced one or more challenges in automating workflows using Actions. Users had issues as the workflows are *hard to test and debug* (14.49%), *too complex for some tasks* (8.7%), and *have inadequate documentation* (7.25%) (Table 6).

> *A typical day of working on GitHub actions is a pain. Having to make your changes, hope it works, commit, push, launch the action, wait 10 min, see it fail, continue. [S-73]*

> *There is a lack of tutorials and guides for GitHub Actions. It's frequently not clear how to achieve a certain configuration, and then it's quite hard and time-consuming to debug. [S-71]*

**Using the marketplace**$_{(25)}$: We asked the users and contributors if they faced any challenges in using the GitHub Marketplace. Seven out of 25 participants stated it is *hard to search for products* and *hard to quality check*. Other than that, one user was not aware of GitHub Marketplace for a long time and felt that the marketplace was *not marketed enough* (Table 5).

> "Marketplace does not provide an effective way to filter and sort based on quality, version, contributions, etc. [S-84]"

> Test-ability is the most frequently reported problem for creating and using Actions. The YAML programming language which is used to develop Action's configuration files was reported as buggy and hard to test. Unfamiliarity with the syntax, not having code IntelliSense, and inadequate documentation about configuring Actions using YAML are the top challenges GitHub users face with configuring workflows. Publishing on the Marketplace is also time-consuming and developers refrain from it.

## 5 DISCUSSION AND RECOMMENDATIONS

**Implication For Researchers:** It only took GitHub Action 18 months to surpass Travis CI, which had dominated the CI scene for over nine years [12]. Many software tools are either being replaced by Actions or being integrated into and packaged as Actions. Researchers can examine how the invention of Actions has changed the ecosystem for software tools. So far, the research community has only looked into the Actions from the users' perspective, i.e. they systematically crawl repositories that are *using* Actions (see Table 1). The field of automated software engineering has been growing extensively. Distinguishing between the providers and adopters is essential since Actions in their nature are atomic building blocks of automation in practice. The actions provide a unique opportunity for researchers to evaluate the tasks that are being automated and the users' feedback, usage, and preferences within practice [36, 37]. In this study, we have highlighted the challenges developers face in automating workflows using Actions, such as Actions being hard to test and debug, missing IntelliSense for configuration files, or having integration and security issues. This also provides new opportunities to explore the field of programming languages, and mining software repositories.

**For the GitHub Actions team:** 21 of our 90 participants do not use Actions, only six of them know that Actions exist, but half of them are unaware of the benefits of Actions. One respondent said:

> "I am not sure of the possibilities. [S-59]"

*Limited resources* is the second-most perceived challenge by Action creators (Table 3). Action users mention *inadequate documentation* and *lack of support* as two of the challenges in automating workflows. Although GitHub has extensive documentation for GitHub Actions[4], it seems many practitioners are either unaware of them or find them unhelpful. We asked developers about the challenges they face in using the Marketplace. One of the responses was:

[4]https://docs.github.com/en/actions

> "The marketplace isn't as relevant, it took me so long to discover it even exists, my developing career would be so easier if I did earlier. [S-11]"

62.32% of the participating developers search Marketplace to identify a proper Action for their tasks. This is while only 21.74% of them state any preference in using the Actions published on the Marketplace. This implied the power of Marketplace as a hub of gathering information but the characteristics of a multi-sided platform are missed from the Marketplace which led to multiple trust and reuse challenges [1, 23]. Implementing stronger mechanisms as such to assist developers in their decision-making is recommended.

Nine out of 25 (36%) developers who did not publish their Action on the marketplace said publishing *takes too much time and effort*. We also recommend GitHub make publishing Actions to the marketplace less strenuous while also ensuring that the process stays effective enough to enforce the best practices. GitHub Marketplace should introduce reviews similar to mobile stores [23–26] that users can benefit from community feedback. To the least, Marketplace should let users filter Actions based on metrics such as the number of contributors, verified creator, etc.

**For Action providers:** Action providers' #1 reason for creating new Actions instead of reusing existing ones is the unavailability of the needed functionality for a particular task. We found out that 39.13% developers create their own Action when they require one while only 2.9% developers check if similar Action was available in other public repositories (Table 6). Also, the second-most common reason why developers switched to a different Action is that they created their own Action. To leverage the re-usability to the fullest extent, we recommend that developers perform a thorough analysis of existing Actions before creating their own. One alternative to creating would be to contribute to an existing Action, which is likely to save time and improve re-usability. We also recommend that developers publish their Actions to the Marketplace so that they get better visibility, more contributors, and improved code quality. Furthermore, the structured status of YAML files and the openness of the Action repositories makes it quite possible to use data analysis tool to mine and compare the functionalities within the existing Actions.

**For Action users:** There are currently over 15,000 Actions across 21 categories in the GitHub Marketplace. It is difficult for potential Action users to decide which Action to use. *Security concerns* is one of the five major developer challenges when automating workflows on GitHub (see Table 6). In order to avoid accidentally using a compromised Action that might have had its code altered and might potentially be used to steal secrets, GitHub suggests referring to reusable Actions using their specific commit SHA[5]. As Decan et al. [9] pointed out, this recommendation is not followed in practice. Our survey showed that the developers who are using an Action prefer *verified creator*, *more stars*, and *more contributors* when they are choosing between Actions. To avoid security issues, GitHub recommends only using actions whose creator can be trusted.

Although Borges et al. [3] warned about relying on the number of stars as a metric for repository quality, stating that more stars could mean effective marketing and advertising strategies and not necessarily solid software engineering principles and practices. However, the number of stars remains a popular metric among practitioners. Therefore, we back our survey finding and recommend potential Action users consider verified creator, more stars, and more contributors when they are choosing between Actions.

[5]https://docs.github.com/en/actions/security-guides/security-hardening-for-github-actions

## 6 THREATS TO VALIDITY

**External Validity:** We used *closed and open invitation* strategies for our survey [22, 32]. Closed invitations allowed us to have more control (such as inviting Action creators). We further sent open invitations via social media channels. To have a better understanding of the participants, we grouped them based on their familiarity with Actions as well as their years of experience. We present our analysis based on 90 complete responses, which should be generalized only in consideration of these demographics and sample size. Nothing in the study specifically limits replication on a bigger group of developers to make our conclusions more robust.

**Construct Validity:** We used thematic analysis for the open-ended questions. The accuracy of this analysis depends on the annotators' accuracy and perception of classification. The exploratory design of the study could have introduced *researcher bias*. To reduce this, annotators worked independently and within iterations. Further, any disagreement was discussed and mediated.

**Internal Validity:** We designed our questionnaire in iterations where the initial draft was tested with a small group for *alternative form reliability* and *content validity*. Also, a later version was reviewed and tested for clarity, bias, and ethics by an independent party. Yet, there is a chance that some survey participants interpret the questions in different ways and find them unclear. Another threat could arise from the arrangement of the options in a ranking-based question which biases participants in choosing top-listed options. We ordered the questions in a natural sequence and in an effort to clarify the context.

**Conclusion Validity:** We draw conclusions in our study based on a survey, which can be inaccurate at times. Surveying software developers do not always provide a comprehensive perspective of real-world practices [28]. However, surveys have been prevalent as a research tool in software engineering. We see the survey as a complementary tool for other types of study such as mining.

## 7 CONCLUSION

This study aims to complement the current literature, which looks beyond the statistics mined from repositories. So far, the Research community has performed mining studies on the usage and adoption of GitHub Actions, where developers integrate an Action into their workflows by customizing a YAML file. To the best of our knowledge, this is the first study that differentiates between Action developers and adopters and further investigates the motivation and decision criteria of the Action adopters. The existing literature focuses on the quantitative analysis of GitHub repositories to investigate the usage of GitHub Actions. Through a qualitative analysis with developers, we present the motivation, decision criteria, and challenges in creating, publishing, and using GitHub Actions. Our results demonstrate the limited leverage of GitHub Marketplace as a multi-sided market. Compared to what is seen in the app stores, here, developers refer to the Marketplace as an aggregator and for searching Actions. Rather, marketplaces should connect users and facilitate feedback loops. Our findings help Action providers identify and improve areas to improve their competitiveness. We believe this study helps researchers understand the status quo and conduct relevant research to facilitate automation practices on software repositories such as GitHub. We found that configuring and debugging the workflow YAML files is one of the most prevalent problems with using Actions. An automatic workflow generator framework can help alleviate this problem and make GitHub Actions more accessible to a larger audience.